\documentclass[conference]{IEEEtran}
\IEEEoverridecommandlockouts
\usepackage{cite}
\usepackage{amsmath,amssymb,amsfonts}
\usepackage{algorithmic}
\usepackage{graphicx}
\usepackage{textcomp}
\usepackage{xcolor}
\def\BibTeX{{\rm B\kern-.05em{\sc i\kern-.025em b}\kern-.08em
    T\kern-.1667em\lower.7ex\hbox{E}\kern-.125emX}}
\begin{document}

\title{Deep Feature Embedding and Hierarchical Classification for Audio Scene Classification}

\author{\IEEEauthorblockN{Lam Pham}
\IEEEauthorblockA{\textit{School of Computing} \\
\textit{University of Kent}\\
Kent, UK \\
ldp7@kent.ac.uk}
\and
\IEEEauthorblockN{Ian McLoughlin}
%\IEEEauthorblockA{\textit{National Engineering Laboratory for Speech and Language Information Processing} \\
\IEEEauthorblockA{\textit{School of Computing} \\
%\textit{University of Science and Technology of China}\\
\textit{University of Kent}\\
%Hefei, China \\
Kent, UK \\
ivm@lintech.org}
\and
\IEEEauthorblockN{Huy Phan}
\IEEEauthorblockA{\textit{School of Computing} \\
\textit{University of Kent}\\
Kent, UK \\
h.phan@kent.ac.uk}
\and
\IEEEauthorblockN{R. Palaniappan}
\IEEEauthorblockA{\textit{School of Computing} \\
\textit{University of Kent}\\
Kent, UK \\
r.palani@kent.ac.uk}
\and
\IEEEauthorblockN{Alfred Mertins}
\IEEEauthorblockA{\textit{Institute for Signal Processing} \\
\textit{University of L\"ubeck}\\
L\"ubeck, Germany \\
alfred.mertins@uni-luebeck.de}

}

\maketitle

\begin{abstract}
In this work, we propose an approach that features deep feature embedding learning and hierarchical classification with triplet loss function for Acoustic Scene Classification (ASC). 
In the one hand, a deep convolutional neural network is firstly trained to learn a feature embedding from scene audio signals. Via the trained convolutional neural network, the learned embedding embeds an input into the embedding feature space and transforms it into a high-level feature vector for representation. In the other hand, in order to exploit the structure of the scene categories, the original scene classification problem is structured into a hierarchy where similar categories are grouped into meta-categories. Then, hierarchical classification is accomplished using deep neural network classifiers associated with triplet loss function. Our experiments show that the proposed system achieves good performance on both the DCASE 2018 Task 1A and 1B datasets, resulting in accuracy gains of $15.6$\% and $16.6$\% absolute over the DCASE 2018 baseline on Task 1A and 1B, respectively. %We also empirically investigate the effects of the employed triplet loss function and the usage of multiple time-frequency inputs in an ensemble fashion. 
\end{abstract}


\section{Introduction}
\label{sec:introduction}
In acoustic scenes, various associated and sporadic event sounds tend to occur within a typical recording. 
We refer to those as foreground sounds, in contrast to background, which is the more constant sound corresponding to that scene. 
Acoustic scene classification (ASC) is complicated by the presence of foreground sounds and by interfering noise, and is characterised by encompassing a very wide range of spectral shapes and temporal sound patterns.
To deal with these challenges, many authors who achieved competitive classification accuracy~\cite{hossein_conv, truc_dca_18, octave_exploring, yuma} on the DCASE 2018 dataset~\cite{dcase2018_db} proposed ensemble models that explore diverse approaches to both input features and learning models.
In particular, Hossein Zeinali \emph{et al.}~\cite{hossein_conv} made use of effective combination of Constant Q Ttransform (CQT) and log-Mel spectrograms.
Firstly, they transferred draw audio into spectrogram, extracting X-vector from these spectrograms. 
Then, they fed these features (both two spectrograms and X-vectors extracted) into one/two-dimensional CNN models.
Eventually, obtained scores were fused to produce the final classification result.
Exploring nearest neighbour filter (NNF), Truc \emph{et al.} \cite{truc_dca_18} extracted NNF spectrogram from log-Mel spectrogram.
Next, the authors fed four spectrograms (coming from from side, average of audio channels and two log-Mel, NNF spectrograms) into separated CNN-based models and fuse four obtained scores.
Deeply focusing on audio channels, Octave Mariotti \emph{et al.}~\cite{octave_exploring} and Yuma et al.~\cite{yuma} experimented on a  wide range of input features (left, right, side and average of channels with log-Mel spectrogram and  Harmonic Percussive Source Separation).
Regarding ensemble models, while Yuma \emph{et al.}~\cite{yuma} proposed a single CNN model similar to VGG configuration, Octave Mariotti \emph{et al.}~\cite{octave_exploring} pursuited an intensive ensemble, evaluating  a variety of deep learning models (VGG8, VGG10, VGG12, Resnet 18, Resnet 34, Resnet 50). 

Another approach relies upon ever more powerful learning models.
For example, Yang \emph{et al.}~\cite{yang_acoustic} proposed a complicated CNN-based architecture called the \emph{xception} network. 
This is inspired by the fact that a deep learning network trained by a wide range of feature scales and over separated channels can result in a very powerful model. 
Indeed, xception achieves the highest score for the DCASE 2018 Task 1A.
Focusing on attention mechanism, an attention-based pooling layer proposed by Zhao Ren \emph{et al.}~\cite{zhaoren_attention} helps to improve the quality of pooling layers compared with traditional pooling layers. 
Exploring different frequency bands in a spectrogram, Phaye \emph{et al.}~\cite{phaye_ica_19} proposed a SubSpectralNet network which is useful to extract discriminative information from 30 sub-spectrograms.
More recently, Hong \emph{et al.}~\cite{hong} proposed a new method that exploits distinct features in sound scenes. 
They firstly applied a deep learning model to extract a bag of similar and distinct features, then leverage this to enforce higher network performance. 
Generally, although the second trend shows complicated network architectures, almost top performances come from ensemble of CNN-based models as mentioned in the first line of methods~\cite{truc_icme,  hossein_conv, truc_dca_18, octave_exploring, yuma}.  

\begin{figure}[t]
	\centering
	\centerline{\includegraphics[width=\linewidth]{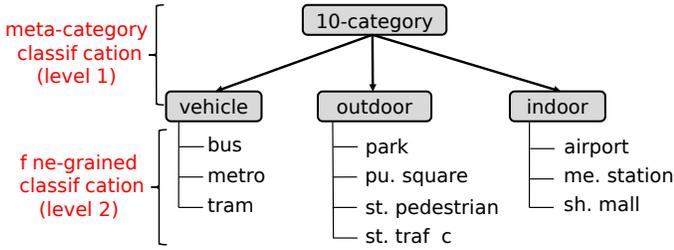}}
	\vspace{-0.2cm}
	\caption{The two-level hierarchy of scene categories constructed based on the categories of the DCASE 2018 datasets.}
	\label{fig:Y1}
\end{figure}

In this paper, we adopt a different approach based on deep feature embedding learning and a hierarchical classification scheme. First, feature embeddings are learned with a deep CNN in a regular classification setting. Rather than using the trained deep CNN for direct classification, it is employed as a feature extractor to embed an audio input into a high-level feature space via the learned embedding. Afterwards, the original ``flat''ASC task, i.e. classification of all categories at once, is structured into multiple hierarchical sub-tasks in a divide-and-conquer manner. In the one hand, the hierarchy is constructed bottom-up. Starting from the original scene categories at the bottom, those categories, that are expected to be acoustically similar, are grouped into a meta-category as demonstrated in Figure \ref{fig:Y1}. The meta-categories, therefore, constitutes the first level of the classification hierarchy. In the other hand, the classification is performed top-down, i.e. classification of the meta-categories is carried out first before classification of categories in a meta-category takes place. The classifiers in the classification hierarchy are realized by deep neural networks (DNNs). Triplet loss function, which was shown to increase Fisher's criterion, is used to trained the DNN classifiers. 

\label{ssec:SHC}
%++++++++++++++
% move to somewhere later
%As using multiple input types has been a rule of thumb in ASC \textcolor{red}{[cite]}, we propose to use three different time-frequency inputs, including log-Mel~\cite{librosa_tool}, Gammatone filter (GAM)~\cite{auditory2009_tool} and Constant Q Transform (CQT)~\cite{librosa_tool}, in the proposed system.
%
\section{The Proposed System}
\label{sec:the_proposed_system}
\subsection{Learning Feature Embeddings}
\label{ssec:proposed_baseline_model}

\begin{figure}[t]
	\centering
	\includegraphics[width=\linewidth]{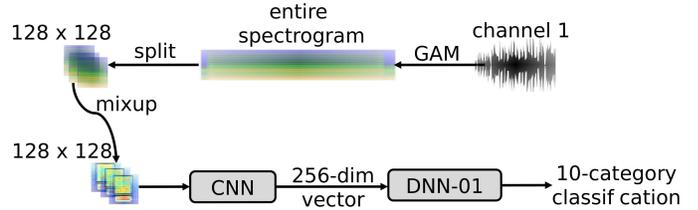}
	\vspace{-0.7cm}
	\caption{Illustration of the processing pipeline to train the CNN for deep feature embedding learning.}
	\label{fig:X1}
\end{figure}

The processing pipeline for deep feature embedding learning using a deep CNN is illustrated in Fig. \ref{fig:X1}. Each acoustic scene signal is firstly transformed into time-frequency image, such as Gammatone spectrogram with 128 Gammatone filters ~\cite{auditory2009_tool}. The time-frequency image is then decomposed into non-overlapping image patches of size $128 \times 128$. Let $\mathbf{X}$ and $\mathbf{y}$ denote an image patch and its one-hot encoding label, respectively. Mixup data augmentation ~\cite{mixup1, mixup2, mixup_image} is then applied on the image patches to generate mixup data: 
\begin{align}
    \mathbf{X}_{\text{mp1}} &= \alpha\mathbf{X}_{1} + (1-\alpha)\mathbf{X}_{2}, \label{eq:mix_up_x1} \\
    \mathbf{X}_{\text{mp2}} &= (1-\alpha)\mathbf{X}_{1} + \alpha\mathbf{X}_{2},     \label{eq:mix_up_x2} \\
    \mathbf{y}_{\text{mp1}} &= \alpha\mathbf{y}_{1} + (1-\alpha)\mathbf{y}_{2}, \label{eq:mix_up_y} \\
    \mathbf{y}_{\text{mp2}} &= (1-\alpha)\mathbf{y}_{1} + \alpha\mathbf{y}_{2}. \label{eq:mix_up_y}
\end{align}

In above equations, $\mathbf{X}_1$ and $\mathbf{X}_2$ are two image patches randomly selected from the set of original image patches with their labels $\mathbf{y}_1$ and $\mathbf{y}_2$, respectively. $\mathbf{X}_{\text{mp1}}$ and $\mathbf{X}_{\text{mp2}}$ are two mixup image patches resulted by mixing $\mathbf{X}_1$ and $\mathbf{X}_2$ with a random mixing coefficient $\alpha$. $\alpha$ is drawn from both uniform distribution and beta distribution. Note that the labels $\mathbf{y}_{\text{mp1}}$ and $\mathbf{y}_{\text{mp2}}$ of the two mixup patches are no longer one-hot labels. 

The resulting mixup data is used to train a network for feature embedding learning. To this end, we propose a deep CNN similar to the VGG network~\cite{vgg_net}. The network architecture and parameters are described in Table \ref{table:CDNN}, comprising Batch Normalization (Bn), Convolutional layers (Cv), Rectified Linear layers (Relu), Average Pooling layers (Ap), Drop-out (Dr) and Fully-Connected Layers (Fl).
%--------------------

%++++++++++++++
%++++++++++++++
\begin{table}[t]
    \caption{The CNN architecture for deep feature embedding learning.} 
        	\vspace{-0.1cm}
    \centering
    \scalebox{0.9}{

    \begin{tabular}{l c} 
        \hline 
            \textbf{Layer}   &  \textbf{Output}  \\
        \hline 
         %Input layer (image patch) & $128{\times}128$          \\
         Bn - Cv ($9{\times}9$) - Relu - Bn - Ap ($2{\times}2$) - Dr (0.1\%)      & $64{\times}64{\times}32$\\
         Bn - Cv ($7{\times}7$) - Relu - Bn - Ap ($2{\times}2$) - Dr (0.1\%)      & $32{\times}32{\times}64$\\
         Bn - Cv ($5{\times}5$) - Relu - Bn - Dr (0.2\%)      & $32{\times}32{\times}128$ \\
         Bn - Cv ($5{\times}5$) - Relu - Bn - Ap ($2{\times}2$) - Dr (0.2\%)       & $16{\times}16{\times}128$\\
         Bn - Cv ($3{\times}3$) - Relu - Bn  - Dr (0.2\%)      & $16{\times}16{\times}256$ \\
         Bn - Cv ($3{\times}3$) - Relu - Bn  - Ap ($2{\times}2$) - Dr (0.2\%)  & $8{\times}8{\times}256$ \\
         Bn - Cv ($8{\times}8$) - Relu - Bn - Dr (0.2\%) & $256$ \\  
         
         \hline 
         Fl - Dr (0.3\%)        &  512         \\
	     Fl - Dr (0.3\%)        &  1024           \\
	     Fl - Dr (0.3\%)        &  10          \\                    
       \hline 
    \end{tabular}
    }
    \label{table:CDNN} 
\end{table}

For clarity, in Fig. \ref{fig:X1} and Table \ref{table:CDNN}, we intentionally separate the deep CNN into two parts: the CNN part for feature learning and the DNN part for classification (denoted as \textbf{DNN-01} to distinguish it from those DNNs in Section \ref{ssec:SHC}). Particularly, instead of using a Global Average Pooling layer at the end of the CNN as other authors do~\cite{Truc_2018, yuma, christian_use}, we design an additional convolutional layer with the kernel size of [$8{\times}8$], that equals to the time-frequency resolution of the output of the previous layer, to capture the interaction across the convolutional channel dimension. Since the labels of the mixup data input are no longer one-hot, we trained the network with Kullback-Leibler (KL) divergence loss rather than the standard cross-entropy loss over all $N$ mixup training image patches:
\begin{align}
    \label{eq:loss_func}
    E_{KL}(\Theta) = \sum_{n=1}^{N}\mathbf{y}_{n}\log(\frac{\mathbf{y}_{n}}{\mathbf{\hat{y}}_{n}})  +  \frac{\lambda}{2}||\Theta||_{2}^{2},
\end{align}
where $\Theta$ denotes the trainable network parameters and $\lambda$ denote the $\ell_2$-norm regularization coefficient. $\mathbf{y_{c}}$ and $\mathbf{\hat{y}_{c}}$  denote the ground-truth and the network output, respectively.

Once the network has been trained, the feature-learnaing CNN part of the network is used as a feature extractor and its last convolutional layer is considered as the deep feature embedding. Presented with a new input, the feature extractor will process the input starting from the first convolutional layer to the embedding layer and produce a high-level feature vector of size 256.

\subsection{Two-level Hierarchical Classification}

%===============================================

%++++++++++++++
%++++++++++++++
\begin{figure}[t]
    \centering
    \includegraphics[width=\linewidth]{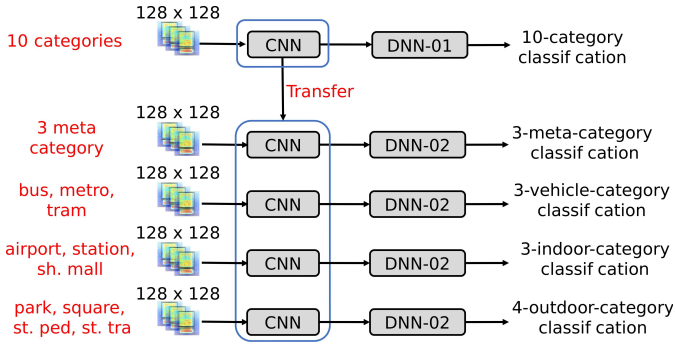}
             \vspace{-0.5cm}
	\caption{Illustration of extracting high-level features from the learned feature embedding to train the DNN classifiers in the hierarchical classification scheme.}
    \label{fig:Y2}
    	\vspace{-0.1cm}
\end{figure}
%++++++++++++++

\begin{table}[t]
	\caption{DNN-02's architecture.} 
	\vspace{-0.1cm}
	\centering
	\begin{tabular}{l c} 
		\hline 
		\textbf{Layer}   &  \textbf{Output Shape}  \\
		\hline 
		Input layer & $256$          \\
		Fl - Dr (0.3\%)        &$512$         \\
		Fl - Dr (0.3\%)        &$1024$           \\
		Fl - Dr (0.3\%)        &$1024$           \\
		Fl - Dr (0.3\%)        &$10$          \\                    
		\hline 
	\end{tabular}
	\label{table:DNN-02} 
	\vspace{-0.1cm}
\end{table}
%-------------------

Most of exiting works follow a ``flat'' classification scheme in which all the scenes categories at classified at once. Differently, we propose to perform the classification hierarchically. The set of scene categories are grouped to form meta-categories. Each meta-category consists of scene categories which are expected to be acoustically similar. In this sense, we construct a two-level hierarchy based on the scene categories in the experimental DCASE 2018 datasets, as shown in Fig. \ref{fig:Y2}. Three meta-categories are formed from 10 scene categories of the DCASE 2018 datasets, including ``vehicle'', ``indoor'', and ``outdoor''. The hierarchical classification is performed in top-down fashion. The meta-categories are classified first, followed by the fine-grained classification of the scene categories in each individual meta-category. As a result, four classifiers are learned: one for meta-category classification (namely mete-category classifier) and three for classification of categories in three meta-categories (namely ``vehicle'' classifier, ``indoor'' classifier, and ``outdoor'' classifier, respectively). An unseen example will be then correctly classified if it is correctly classified by the classifiers at both levels. For example, a ``bus'' scene example is correctly classified if it is both correctly classified as ``vehicle'' by the meta-category classifier and as ``bus'' by the ``vehicle'' classifier. A misclassifcation by one of the classifiers will result in the example is wrongly classified.  

The classifiers involving in the hierarchical classification are realized by DNNs, denoted as \textbf{DNN-02}s. Via the learned embedding presented in Section \ref{sec:the_proposed_system}, 256-dimensional high-level feature vectors are obtained for the mixup image patches and used to train the \textbf{DNN-02}s. In doing this, we effectively transfer the CNN part of the trained CNN in Section \ref{sec:the_proposed_system}, freeze its parameters, and use it as a feature extractor before presenting the extracted features to a \textbf{DNN-02}, as illustrated in Fig. \ref{fig:Y2}. Note that the \textbf{DNN-02}s share a common architecture but are trained separately depending on the sub-tasks in the hierarchical classification. Each \textbf{DNN-02} comprises four fully-connected layers and parametrized as in Table \ref{table:DNN-02}.
%--------------------

%=============== described in experiments====================
%To evaluate effectiveness of transferred learning technique recently mentioned, we do further experiments using four proposed baseline models (C-DNN baseline mentioned in Section \ref{ssec:proposed_baseline_model}). 
%While the first baseline model is used to classify three groups of vehicle, indoor and outdoor places, three remaining models classify classes in each group
%We call this approach as four-CDNNs hierarchical scheme and compare it with the SHC.
%By setting four-CDNNs hierarchical scheme, every model in this scheme is challenged with fewer classes. 
%It shows 3 for three groups at the first level, 3 for classes in indoor/vehicle and 4 for outdoor at the second level. 
%As regards input feature, while four-CDNNs hierarchical scheme receives image patches from front-end feature extraction, SHC receives high-level feature extracted from C-DNN baseline (noting that C-DNN baseline used  to extract feature for SHC is challenge with all 10 classes).
% 
%\subsection{Triplet Loss Applied for Hierarchical Classification Scheme}
%\label{ssec:triplet}
%

In addition to the KL-divergence loss, we additionally employ triplet loss function~\cite{tripletloss} to train the \textbf{DNN-02}s to encourage the networks to improve its discrimination power. Triplet loss function has been shown to be efficient to learn a metric to minimize same-category distances and maximize between-category distances simultaneously, and hence, enhance the Fisher's criterion. Supposed that we present two samples of different categories to a \textbf{DNN-02}, and denote the ground-truth of the first sample as the anchor $\mathbf{a}$, the prediction for the first sample as positive $\mathbf{p}$, and the prediction for the second sample as positive $\mathbf{n}$, the triplet loss is given as
\begin{align}
    \label{eq:triplet_loss}
    E_{triplet} = \max(d(\mathbf{a},\mathbf{p}) - d(\mathbf{a},\mathbf{n}) + margin, 0), 
\end{align}
where \(d\) is squared Euclidean distance and the \(margin\) is set to $0.3$.

The final loss function is, therefore, a combination of the KL-divergence loss and the triplet loss:
\begin{equation}
    \label{eq:final_loss}
    E(\Theta) = \gamma E_{KL}(\Theta) + (1-\gamma)E_{triplet}(\Theta),
\end{equation}
where \(E_{KL}\) is the KL-divergence loss given in (\ref{eq:loss_func}). %The coefficient \(\gamma\) is experimentally set to $0.2$. 
%((\(\gamma\) is selected after evaluating from 0.1 to 0.9 with the step of 0.1).)

\begin{table}[b]
	\caption{The number of scene recordings corresponding to each scene categories in the training set (Train. set) and evaluation set (Eval. set) of the DCASE 2018 Task 1A \& 1B development datasets~\cite{dcase2018_db}.} 
	\vspace{-0.1cm}
	\centering
	\begin{tabular}{l c c c c} 
		\hline 
		\textbf{Category}         & \textbf{Task 1A}        & \textbf{Task 1A}      & \textbf{Task 1B}        & \textbf{Task 1B}     \\ 
		& \textbf{Train. set}       & \textbf{Eval. set}     & \textbf{Train. set}       & \textbf{Eval. set} \\ 
		\hline 
		Airport 	       & $599$                   & $265$                 & $707$                   & $301$ \\        
		Bus     	       & $622$                   & $242$                 & $730$                   & $278$ \\        
		Metro	 	       & $603$                   & $261$                 & $711$                   & $297$ \\        
		Metro Stattion 	       & $605$                   & $259$                 & $713$                   & $295$ \\        
		Park                   & $622$                   & $243$                 & $730$                   & $278$ \\        
		Public Square          & $648$                   & $216$                 & $756$                   & $252$ \\        
		Shopping Mall 	       & $585$                   & $279$                 & $693$                   & $315$ \\        
		Street Pedestrian      & $617$                   & $247$                 & $725$                   & $283$ \\        
		Street Traffic 	       & $618$                   & $246$                 & $726$                   & $282$ \\        
		Tram 	               & $603$                   & $261$                 & $711$                   & $297$ \\        
		\hline 
	\end{tabular}    
	\label{table:dataset} 
\end{table}
\subsection{Ensemble with Multiple Time-Frequency Inputs}
\label{ssec:ensemble}
Using multiple input types has been a rule of thumb in ASC ~\cite{lam, lam_int}. We, therefore, propose to use three different time-frequency inputs, including log-Mel~\cite{librosa_tool}, Gammatone filter (GAM)~\cite{auditory2009_tool}, and Constant Q Transform (CQT)~\cite{librosa_tool}, to form an ensemble of three systems. 
The final decision of each classification task (meta-category classification at the level 1 or fine-grained classifications at the level 2 shown in Figure \ref{fig:Y1}) is obtained by aggregating the individual decisions of the three classifiers (each with one type of spectrogram)  in the ensemble and the final classification label is determined via maximum posterior probability: 
\begin{align}
    \label{eq:final_res}
    \hat{y} = arg max (\mathbf{\bar{p}}_{\text{log-Mel}} + \mathbf{\bar{p}}_{\text{GAM}} + \mathbf{\bar{p}}_{\text{CQT}}),
\end{align}
where $\mathbf{\bar{p}}$ denotes the posterior probability output of a classification model and $\hat{y}$ denotes the final label.

\begin{figure*}[t]
	\centering
	\includegraphics[width=0.9\linewidth, height=5cm]{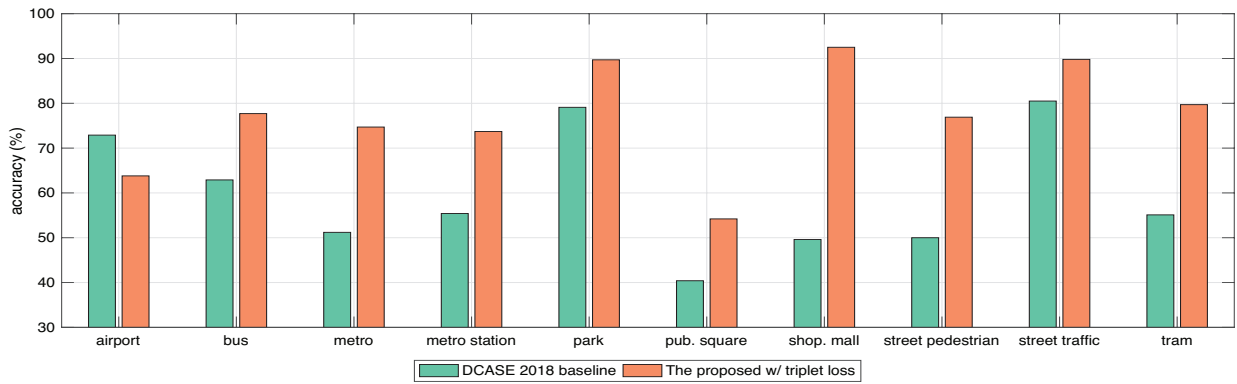}
	%\vspace{-1.0cm}
	\caption{Category-wise performance comparison between the proposed system with triplet loss and the DCASE 2018 baseline on \textbf{Task 1A}.}
	\label{fig:Z15}
\end{figure*}
%++++++++++++++
\begin{table}[t]
	\caption{Performance comparison between the proposed systems, the DCASE 2018 baseline, and the developed baseline.} 
	\vspace{-0.1cm}
	\centering
	\begin{tabular}{l c c} 
		\hline 
		\textbf{System}                   & \textbf{Task 1A}  & \textbf{Task 1B}  \\ [0.5ex] 
		\hline 
		DCASE 2018 baseline ~\cite{dcase2018_db}      & $59.7$  & $45.6$ \\
		The developed baseline                                & $70.9$  & $61.1$ \\	    
		The proposed w/o triplet loss                                                     & $73.3$ &  $\textbf{62.2}$ \\
		The proposed w/ triplet loss                             & $\textbf{75.3}$   & $58.9$ \\        
		\hline 
	\end{tabular}    
	\label{table:c_base} 
\end{table}

\section{Experiments}
\label{sec:result}
\subsection{DCASE 2018 Datasets}
\label{ssec:data}
%------------------------------------------------------------------------------------------

%---------------------------------------------------------------------------------
%

%------------------------------------------------------------------------------------------
%++++++++++++++

Our experiments were based on the DCASE 2018 Task 1A and 1B development datasets~\cite{dcase2018_db}. 
The audio signals in Task 1A  was recorded at a sample rate of $44.1$\,kHz by only one device (known as device A) with $10$-second long for each recording. 
For Task 1B, all recordings using the device A from Task 1A are reused.
In addition, new recordings with two different devices (device B \& device C), were added (72 recordings from each device for every category). The goal of Task 1B is to evaluate the performance on the device B and C when there are mismatched devices in real-world applications. It should be noted the imbalance of Task 1B data as there was only $4$ hours of data recorded with the devices B \& C compared with $24$ hours of data recorded with the device A. Adhering to the setting of DCASE 2018 challenge, we divided the development dataset into a training and evaluation subsets (Train. set and Eval. set) as shown in Table \ref{table:dataset}. 
%For Task 1B, we evaluated performance on the device B \& C data as in the challenge.% (only 36 segments recorded over device B \& C in Eva. set in Task 1B).

\subsection{Baselines}
\label{ssec:baseline}

Besides comparison with the DCASE 2018 baseline and the results reported in previous works, we used the CNN used for deep feature embedding learning in Section \ref{ssec:proposed_baseline_model} as the developed baseline to justify the impact of the learned deep feature embedding and the hierarchical classification scheme. When being used as a classification baseline, the CNN was trained to classify 10 categories of the datasets as in typical setting.

%% THIS IS THE DESCRIPTION OF THE BASELINE WITH FOUR CNNs. I LEAVE IT HERE FOR REUSE IF NECESSARY  
%Using the same classifcation hierarchy as the proposed system, the baseline, however, consists of four CNN classifiers that share a similar architecture with the CNN used for learning feature embedding in Section \ref{ssec:proposed_baseline_model}. There are two main differences between the baseline and the proposed system. First, the CNN classifiers of the baseline were trained using the time-frequency image inputs while the DNN classifiers of the proposed system were trained with the the high-level features extracted via the learned embedding. Second, the CNN classifiers of the baseline were trained to minimize the KL-divergence loss whereas a combination of the KL-divergence loss and the triplet loss was used to train the DNN classifiers of the proposed system.

\subsection{Other parameters}
\label{ssec:parameters}

The time-frequency image features, i.e. Gammatone, log-Mel, and CQT spectrogram, were obtained via a short-time window size of 43\,ms and hop size of 6\,ms. All of them have a common number of filter of 128. 

The networks were implemented using the Tensorflow framework. The coefficient $\lambda$ in (\ref{eq:loss_func}) was set to $10^{-4}$, and $\gamma$ in (\ref{eq:final_loss}) was experimentally set to $0.2$. The network training was accomplished with Adam optimizer~\cite{kingma2014adam} with the learning rate of $10^{-4}$, a batch size of $100$, and stop after 100 epoches.

\subsection{Experimental Results}
%++++++++++++++

Performance obtained by the proposed system, the developed baseline, and the DCASE 2018 baseline are shown in Table \ref{table:c_base}. As can be seen, the propose system outperforms all the DCASE 2018 baseline with a large margin, $15.6$\% absolute (with triplet loss) on Task 1A and $16.6$\% absolute on Task 1B (without triplet loss). Improvements on individual categories can also be seen, as shown in Fig.~\ref{fig:Z15} for a comparison between the proposed system with triplet loss and the DCASE 2018 baseline on Task 1A, with several categories enjoying a significant gain of more than $20\%$, such as ``shopping mall'', ``tram'', ``metro'', ``street-pedestrian''. 

\begin{figure}[t]
	\centering
	\includegraphics[width=1\linewidth]{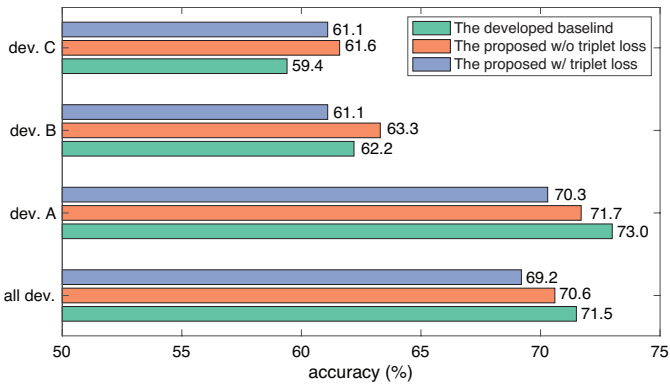}
	%\vspace{-1cm}
	\caption{Accuracy obtained by the systems developed in this work on different devices of \textbf{Task 1B}.}
	\label{fig:Z13}
\end{figure}

Compared to the developed baseline, the proposed system leads to an accuracy gain of $2.4\%$ and $1.1\%$ on Task 1A and Task 1B, respectively, when the triplet loss is not used. When the triplet loss is used, a significant accuracy improvement is seen on Task 1A: $2.4\%$ absolute compared to that without triplet loss and $4.4\%$ compared to the developed baseline thanks to the proposed hierarchical classification scheme. However, using triplet loss seems to be counter-productive on Task 1B as the accuracy is reduced by $3.3\%$ absolute in comparison to the system without  triplet loss. This is presumably due to the device mismatch or the lack of training data on the target devices (device B \& C) or both. However, average over all the devices, the proposed system with triplet loss outperforms all other counterparts, as shown in Fig. \ref{fig:Z13}. 

We further collate the results reported in previous works (both the DCASE 2018 challenge submission systems and the recent works) and provide a comprehensive performance comparison on Task 1A and Task 1B in Tables \ref{table:c_1a_state} and \ref{table:c_1b_state}, respectively. It should be noted that there are inconsistencies between the accuracies reported in the DCASE 2018 technical reports and those published in DCASE 2018 challenge website~\footnote{http://dcase.community/challenge2018/}. The results in Tables \ref{table:c_1a_state} and \ref{table:c_1b_state} are collated from the technical reports which are the original sources of the reported accuracies. For clarity, we only cover top 10 DCASE 2018 challenge submissions in the tables. In the one hand, the proposed system outperforms the recent works (i.e. after the DCASE 2018 challenge) on Task 1A  while retaining as top-3 performer in the context of the DCASE 2018 submission systems. In the other hand, our proposed system achieves state-of-the-art results on Task 1B, achieving an accuracy of $66.9\%$ and outperforming both the DCASE 2018 submission systems and the previous works. 

%++++++++++++++

\subsection{Discussion}
%++++++++++++++

\begin{figure*}[t]
	\centering
	\includegraphics[width=1\linewidth]{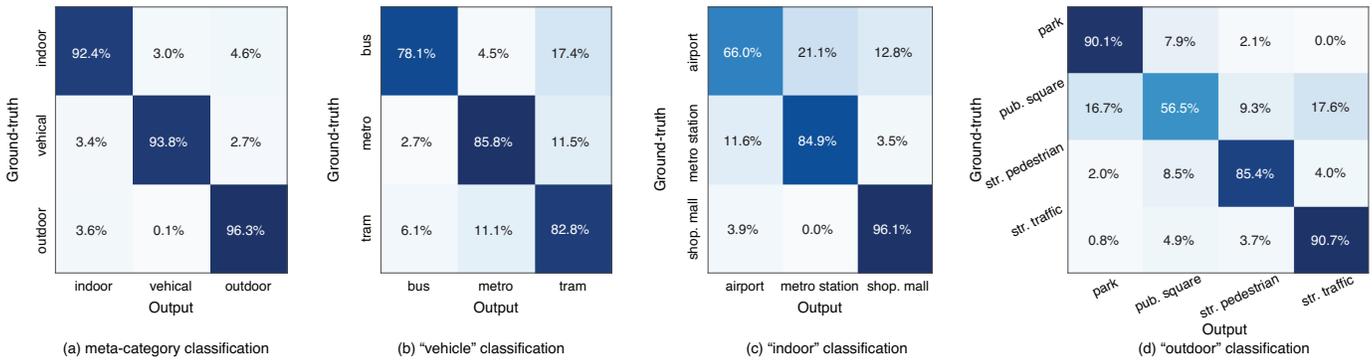}
	%\vspace{-1cm}
	\caption{Confusion matrices obtained by different classifiers in the proposed hierarchical classification scheme on \textbf{Task 1A}.}
	\label{fig:Z11}
\end{figure*}

To shed light on the performance of the classifiers in the proposed hierarchical classification scheme, we shown their confusion matrices in Fig.~\ref{fig:Z11}. Overall, the meta-categories can be discriminated very well with an average accuracy of $94$\% achieved by the meta-category classifier. Given the good performance of the meta-category classifier, the test examples are expected to be directed to the correct groups in the lower level. Even though the fine-grained classifiers' performance are not as good as that of the meta-category classifier since the categories in a group tend to be similar acoustically, they are expected to perform better than the case of ``flat'' classification with 10 classes at once. The reason is, in one group, the classification subtask is able to avoid the confusion between its categories and those in other groups. 

\begin{figure}[t]
	\centering
	\includegraphics[width=1\linewidth]{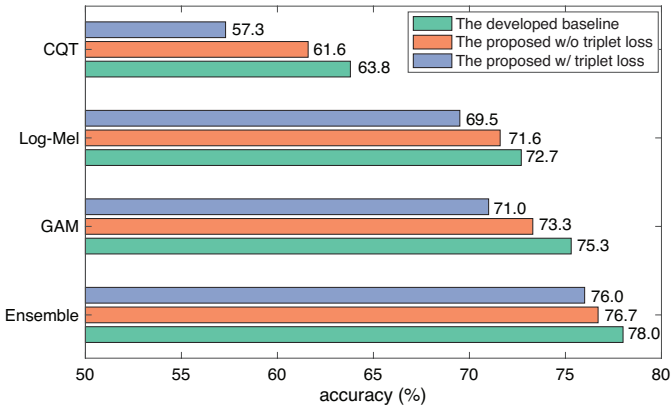}
	% \vspace{-1cm}
	\caption{Performance of individual time-frequency representations and their ensemble on \textbf{Task 1A}.}
	\label{fig:Z12}
\end{figure}

Overall, out of the individual time-frequency inputs (i.e. Gammatone spectrogram, log-Mel spectrogram, and CQT spectrogram), Gammatone spectrogram seems to perform best as shown in Fig. \ref{fig:Z12} while CQT spectrogram  is the worst. However, aggregation the classification outputs of all three results in significant improvements over the individual ones. This is observed over all systems, the proposed system with triplet loss, the proposed system without triplet loss, and the developed baseline. It is expected as different time-frequency representations have been shown to be good for different scene categories, and their individual strength is leveraged in the ensemble to bring up performance gain.

\begin{table}[t]
	\caption{Comparison between DCASE2018 baseline, the top-10 DCASE 2018 challenge (top), recent papers (middle), and the proposed system (bottom) on \textbf{Task 1A}.} 
	\vspace{-0.1cm}
	\centering
	\begin{tabular}{l l c} 
		\hline 
		\textbf{System}                                         &\textbf{Method}    &\textbf{Acc. (\%)}        \\ [0.5ex] 
		
		\hline 
		DCASE2018 Baseline~\cite{dc_2018_bsl}                  &CNN    & $59.7$    \\   
		Li~\cite{dc_2018_t10}                                  &DNN-biLSTM    & $72.9$    \\     
		Jung~\cite{jung_dca_18}                                &Ens. of CNN-SVM    & $73.5$    \\     
		Hao~\cite{dc_2018_t08}                                 &Ens. of biLSTM-CNN    & $73.6$    \\     
		Christian~\cite{dc_2018_t07}                           &CNN-Voting    & $74.7$    \\     
		Zhang~\cite{dc_2018_t06}                               &CNN-SVM    & $75.3$    \\     
		Li~\cite{dc_2018_t05}                                  &Ens. of CNN, DNN    & $76.6$    \\     
		Dang~\cite{dc_2018_t04}                                &Ens. of CNNs   & $76.7$    \\     
		Yuma~\cite{yuma}                                       &Ens. of CNNs    & $76.9$    \\     
		Octave~\cite{octave_exploring}                         &Ens. of CNNs     & $79.3$    \\    %
		Yang~\cite{yang_acoustic}                              &Xception CNN    & $\textbf{79.8}$    \\    % 
		
		\hline 
		Bai~\cite{bai}                                         &Hybrid-DNN             & $66.1$   \\    
		Zhao~\cite{zhao_ica_19}                                &CNN                    & $72.6$   \\   
		Phaye~\cite{phaye_ica_19}                              &SubSpectralNet CNN     & $74.1$   \\
		Zeinali~\cite{hossein_conv}                            &Ens. of CNNs       & $77.5$   \\            
		
		\hline 
		%\textbf{Proposed baseline}                             &CNN                     & $70.9$     \\
		%\textbf{SHC}                                           &CNNs               & $73.3$     \\
		%\textbf{SHC wi. Triplet loss}                          &CNNs               & $75.3$     \\
		%\textbf{Proposed baseline wi. Ens.}             &Ens. of CNNs                    & $76.0$     \\
		\textbf{The proposed  w/ triplet loss}                &Ens. of hier. DNNs   & $78.0$     \\
		\hline 
	\end{tabular}    
	\label{table:c_1a_state} 
\end{table}
%------------------------------------------------------------------------------------------
%------------------------------------------------------------------------------------------
\begin{table}[t]
	\caption{Comparison between the DCASE 2018 baseline, the top-7 DCASE 2018 challenge (top), the recent papers (middle), and the proposed system (bottom) on \textbf{Task 1B} (only devices B \& C).}
	\vspace{-0.1cm}
	\centering
	\begin{tabular}{l l c} 
		\hline 
		\textbf{System}                       &\textbf{Method}                    &\textbf{Acc. (\%)}        \\ [0.5ex] 
		
		\hline 
		DCASE2018 Baseline~\cite{dc_2018_bsl}     &CNN                       &$45.6$ \\   
		Li~\cite{dc_2018_tb07}                    &Ens. of CNN, DNN      &$51.7$ \\
		Tchorz~\cite{dc_2018_tb06}                &LSTM                      &$53.9$ \\ 
		Kong~\cite{dc_2018_tb05}                  &CNN                       &$57.5$ \\ 
		Wang~\cite{dc_2018_tb04}                  &Self-attention CNN        &$57.5$ \\ 
		Waldekar~\cite{dc_2018_tb03}              &Ens. of CNNs          &$57.8$ \\ 
		Zhao~\cite{zhao_dca_18}                   &CNN                       &$58.3$ \\
		Truc~\cite{truc_dca_18}                   &Ens. of CNNs          &$63.6$ \\
		
		\hline 
		Zhao~\cite{zhao_ica_19}                   &CNN                       & $63.3$  \\   
		Truc~\cite{truc_int}                      &CNN, Mix. of Experts   & $64.7$  \\          
		Yang~\cite{dc_2018_yang}                  &Xception CNN              & $65.1$  \\          
		Truc~\cite{truc_icme}                     &Ens. of CNNs          & $66.1$ \\
		
		\hline 
		%\textbf{Proposed baseline}               &CNN                        & $61.1$     \\
		%\textbf{SHC}                             &CNNs                  & $62.2$     \\
		%\textbf{SHC wi. Triplet loss}            &CNNs                  & $58.9$     \\
		%\textbf{Proposed baseline wi. Ens.}                    &Ens. of CNNs      & 65.2    \\
		\textbf{The proposed w/o triplet loss}                    &Ens. of hier. DNNs      & $\textbf{66.9}$     \\
		\hline 
	\end{tabular}    
	\label{table:c_1b_state} 
\end{table}
%------------------------------------------------------------------------------------------

%++++++++++++++
%++++++++++++++
%------------------------------------------------------------------------------------------

\section{Conclusion}
We have presented an approach that learns deep feature embedding to extract high-level features for audio scene signals via a deep CNN and proposed a novel hierarchical classification scheme to accomplish the scene classification. In the classification hierarchy, the similar scene categories are grouped into meta-categories. Meta-category classification was carried out first, followed by the fine-grained classification in the groups. DNNs were trained with triplet loss to play the role of the classifiers in the classification hierarchy. Experiments on the DCASE 2018 Task 1A and 1B datasets demonstrated that the proposed methods significantly outperformed the DCASE 2018 baseline while achieving highly competitive results compared to state-of-the-art systems. In future work, it is worth further experimenting with deeper-level hierarchical schemes with large number of categories as well as with data-driven clustering approaches.

%\section{REFERENCES}
%\label{sec:refs}

\bibliographystyle{IEEEbib}
\bibliography{strings,refs}

\end{document}